# Tailoring Hot Exciton Dynamics in 2D Hybrid Perovskites through Cation Modification

Daniel B. Straus[1,#], Sebastian Hurtado-Parra[2], Natasha Iotov[2], Qinghua Zhao[1], Michael R. Gau[1], Patrick J. Carroll[1], James M. Kikkawa[2], and Cherie R. Kagan*[1,3,4]

Departments of [1]Chemistry, [2]Physics, [3]Electrical and Systems Engineering and [4]Materials Science and Engineering, University of Pennsylvania, Philadelphia, PA 19130

[#]Present address: Department of Chemistry, Princeton University, Princeton, NJ 08544

[*]Author to whom correspondence should be addressed. Email: kagan@seas.upenn.edu

**Abstract:**

We report a family of two-dimensional hybrid perovskites (2DHPs) based on phenethylammonium lead iodide ((PEA)$_2$PbI$_4$) that show complex structure in their low-temperature excitonic absorption and photoluminescence (PL) spectra as well as hot exciton PL. We replace the 2-position (*ortho*) H on the phenyl group of the PEA cation with F, Cl, or Br to systematically increase the cation's cross-sectional area and mass and study changes in the excitonic structure. These single atom substitutions substantially change the observable number of and spacing between discrete resonances in the excitonic absorption and PL spectra and drastically increase the amount of hot exciton PL that violates Kasha's rule by over an order of magnitude. To fit the progressively larger cations, the inorganic framework distorts and is strained, reducing the Pb-I-Pb bond angles and increasing the 2DHP band gap. Correlation between the 2DHP structure and steady-state and time-resolved spectra suggests the complex structure of resonances arises from one or two manifolds of states, depending on the 2DHP Pb-I-Pb bond angle (as)symmetry, and the resonances within a manifold are regularly spaced with an energy separation that decreases as the mass of the cation increases. The uniform separation between resonances and the dynamics that show excitons can only relax to the next-lowest state are consistent with a vibronic progression caused by a vibrational mode on the cation. These results demonstrate that simple changes to the structure of the cation can be used to tailor the properties and dynamics of the confined excitons without directly modifying the inorganic framework.

TOC Graphic

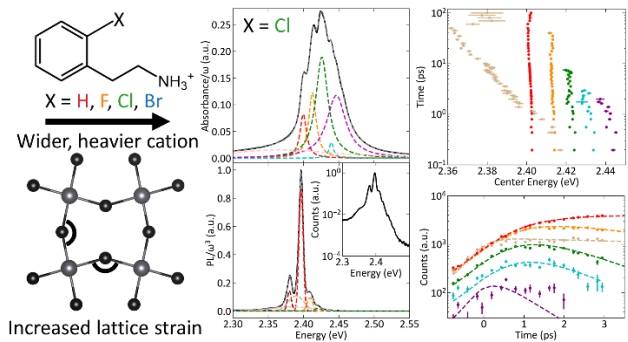

An emerging class of two-dimensional semiconductors is 2D organic-inorganic metal-halide hybrid perovskites (2DHPs).[1–3] 2DHPs are solution-processable materials that spontaneously crystallize upon mixing of their precursors into sheets of corner-sharing metal-halide octahedra separated by organoammonium cations. Because of their stoichiometric nature, 2DHPs are often considered to be "perfect" quantum well superlattices, free from interfacial roughness and defects typically seen in 2D materials grown by molecular beam epitaxy (MBE).[3] The optical and electronic properties of 2DHPs depend on composition and can be tailored through the choice of metal, halide, and cation. For example, lead halide 2DHPs have band gaps tunable between 2 and 4 eV and form heterostructures, akin to those in MBE-grown superlattices, with both Type I and Type II band alignments. In Type I 2DHPs, the alternating organic and inorganic layers result in strong quantum and dielectric confinement effects that localize excitons to the inorganic framework with binding energies >150 meV.[3]

In contrast to Type I MBE-grown 2D materials,[4] the wider band gap layers of intervening organic cations in 2DHPs directly interact with strongly confined excitons.[5,6] In our previous work, we studied the 2DHP phenethylammonium lead iodide (($PEA)_2PbI_4$; $PEA=(C_6H_5C_2H_4NH_3^+)$). 99.7% of the excitonic photoluminescence (PL) is emitted out of the lowest-lying excited state at 15 K. However, 0.3% of the PL is emitted out of discrete higher energy resonances that also appear at similar energies in the absorption spectrum, each separated by ~40 meV from the lowest-lying resonance and from one another.[5] The existence of hot PL violates Kasha's rule, which states carriers relax to the lowest-lying excited state before fluorescing.[7] PL from these higher-energy resonances competes with relaxation to the lowest-lying excited state, and the kinetics combined with the regular spacing of the resonances led us to hypothesize that we observed a phonon progression. Importantly, theoretical calculations supported the presence of phonons located exclusively on the organic cation with energy near 40 meV. Our hypothesis that the excitonic structure is caused by a vibronic progression has been controversial, though all proposed explanations we are aware of involve some form of exciton-phonon coupling.[8–11] Coupling of excitons to phonons on the organic and inorganic frameworks has been optically observed in 2DHPs[6,12,13] and in all-inorganic 3D halide perovskite nanocrystals.[14]

Here, we demonstrate that exciton dynamics in 2DHPs can be modified by tuning the structure of the organic cation, even though the band edges in 2DHPs are formed by Pb and I orbitals.[15] We modify the PEA cation by introducing a halogen atom in the 2-position of the phenyl group (*ortho* to ethylammonium), forming the perovskites 2-fluorophenethylammonium lead iodide (2-F), 2-chlorophenethylammonium lead iodide (2-Cl), and 2-bromophenethylammonium lead iodide (2-Br). These single atomic substitutions substantially change the energy of, observable number of, and spacing between the discrete resonances in the excitonic absorption and PL spectra and drastically increase the amount of hot exciton PL that violates Kasha's rule with over an order of magnitude greater hot exciton PL in 2-F and 2-Cl than in $(PEA)_2PbI_4$. These results support our original hypothesis that the excitonic structure is caused by a vibronic progression. Our findings have profound implications on the design and application of 2D perovskites because the cation can be used to finely tune the properties and dynamics of the confined excitons without the need to change the composition of the inorganic framework.

**Results and Discussion**

We synthesize $(PEA)_2PbI_4$, 2-F, 2-Cl, and 2-Br single crystals and structurally characterize the crystals by X-ray diffraction measurements at 100 K. The crystal structures are detailed in Figure S1 and Table S1. Figure 1A-E illustrates the structural differences between the Pb-I frameworks of (PEA)$_2$PbI$_4$, 2-F, 2-Cl, and 2-Br, with the Pb-I-Pb bond angles plotted in stars in Figure 1F. In contrast to cubic perovskites which have Pb-I-Pb bond angles of 180° (Figure 1A), the inorganic frameworks in the 2DHPs discussed herein are distorted with Pb-I-Pb bond angles <180°. Introducing larger atoms in the 2-position of the PEA cation increases their cross-sectional area, which creates additional distortion and strain on the inorganic framework[3,16] (Figures 1B-E) and also increases the distance between Pb-I layers. The 2-Br cation is so large that it leads to out-of-plane buckling of the Pb-I framework (blue, Figure 1E) that is not present in (PEA)$_2$PbI$_4$, 2-F, or 2-Cl.[16] Importantly, 2-Cl has symmetric Pb-I-Pb bond angles (Figure 1D), whereas (PEA)$_2$PbI$_4$, 2-F, and 2-Br exhibit two distinct Pb-I-Pb bond angles (Figure 1B-C, F). 2-F also exhibits additional disorder in the organic framework compared to (PEA)$_2$PbI$_4$, 2-Cl, or 2-Br with twice as many possible cation orientations as 2-Cl despite being part of the same space group (Figure S2).

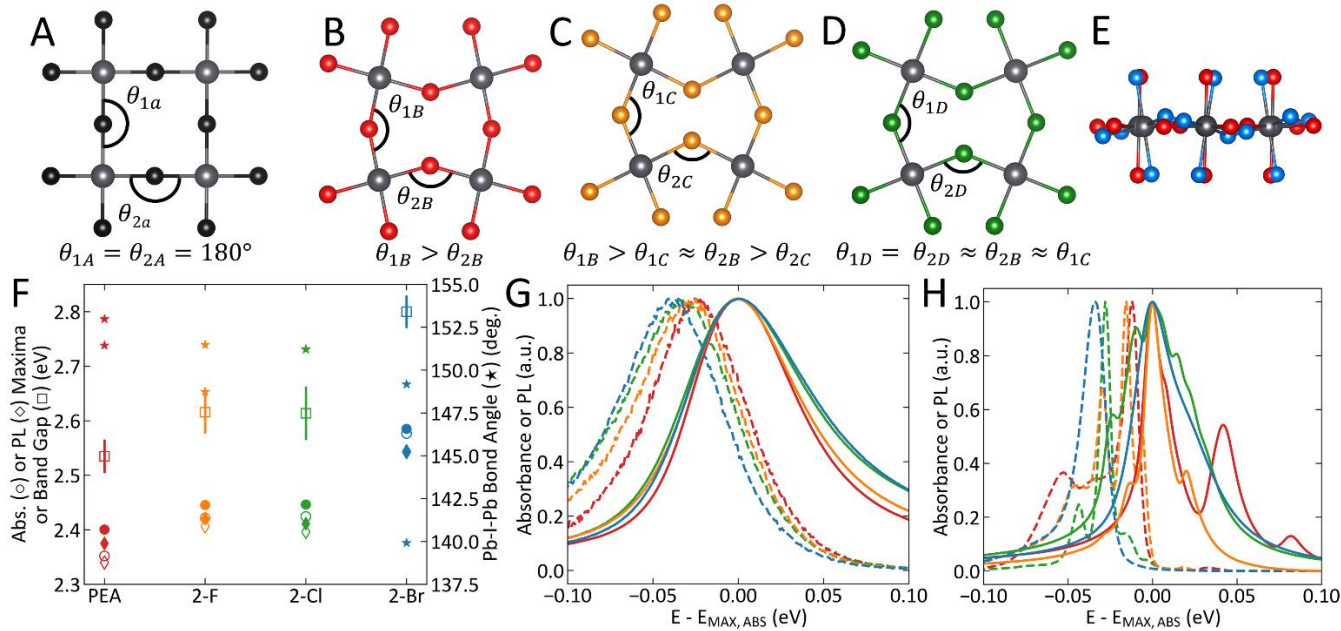

Figure 1: Representation of the in-plane metal-halide structure for (A) an ideal cubic perovskite lattice, (B) (PEA)$_2$PbI$_4$, (C) 2-F, and (D) 2-Cl. (E) Projection down an inorganic sheet for 2-Br (blue) and (PEA)$_2$PbI$_4$ (red) showing corrugation in 2-Br. (F) Comparison of the Pb-I-Pb bond angles $\theta_1$ and $\theta_2$ (stars), excitonic absorption (circles) and PL (diamonds) maxima at 300 K (filled symbols) and 10 K (open symbols), and band gaps at 10 K (open squares). 300 K (G) and 10 K (H) excitonic absorption (solid) and PL (dashed) spectra of (PEA)$_2$PbI$_4$ (red), 2-F (orange), 2-Cl (red), and 2-Br (blue) using 3.10 eV excitation. Energy zero of spectra in (G) and (H) is set to the absorption maximum for each material.

We perform spectroscopic studies on thin films with thicknesses of 27-32 nm of (PEA)$_2$PbI$_4$, 2-F, 2-Cl, and 2-Br spin-cast out of acetonitrile. Room temperature powder diffraction patterns of thin films of (PEA)$_2$PbI$_4$, 2-F, 2-Cl, and 2-Br on sapphire substrates (Figure S3) all exhibit regularly spaced reflections indicative of the layered structure being parallel to the substrate, and the trend of increasing interlayer spacing for larger cations matches that found in the 100 K crystal structures.

Figure 1F shows that smaller Pb-I-Pb bond angles result in blue-shifted absorption (circles) and PL (diamonds) spectra and therefore band gaps (squares, Figure S4, and Table S2), consistent with tight-binding calculations in Sn-I 2DHPs,[16] at both 300 K (filled symbols) and 10 K (open symbols). The energy of the absorption and PL maxima correlates with the larger of the bond angles, so 2-F and 2-Cl have similar absorption and PL maxima despite the additional distortion along one direction in 2-F. Figure 1G shows room temperature excitonic absorption (solid) and PL (dashed) spectra for $(PEA)_2PbI_4$ (red), 2-F (orange), 2-Cl (green), and 2-Br (blue) centered at the absorption maximum. Previously, we showed that increasing the interlayer spacing between inorganic layers by changing the length of the cation results in a more disordered energy landscape, broadening the excitonic absorption.[6] Here, we see the same trend at 300 K as the width of the excitonic absorption progressively increases for 2-F, 2-Cl, and 2-Br compared to $(PEA)_2PbI_4$. 300 K PL spectra are also consistent with our previous report. The PL spectra are nearly invariant in width but have an increased Stokes shift as the cation becomes larger,[6] indicating that excitons find energetic minima before recombining, consistent with disorder and/or polaron formation.[5,9,13,17–21]

As the temperature is decreased from 300 K to 10 K, the excitonic absorption and PL resonances for 2-F, 2-Cl, and 2-Br linearly red-shift (Figure S5-S6). The excitonic absorption maximum as a function of temperature $(E_{MAX,ABS}(T))$ is well-fit by a linear model using the equation

$$E_{MAX,ABS}(T) = E_{MAX,ABS}(0) - \alpha T.$$

The excitonic absorption resonance can be used as a proxy to evaluate the temperature dependence of the band gap in 2DHPs because the degree of quantum and dielectric confinement should not change significantly with temperature.[6] For 2-F, $\alpha$=-0.080(2) meV/K, and similarly for 2-Cl, $\alpha$ = -0.076(2) meV/K. 2-Br shows smaller changes in the absorption maxima with temperature, and α=-0.032(3) meV/K. The negative values of α are consistent with previous reports by us and others for Pb-based hybrid perovskites[6,22,23] as well as lead chalcogenides.[24–26] α is much smaller than that we observed for $(PEA)_2PbI_4$ and its derivatives which have longer cations with unchanged cross-sectional area, where α ranged from -0.14 to -0.16 meV/K,[27] indicative of a weaker temperature dependence of the electronic properties of 2DHPs as the Pb-I-Pb bond angle decreases and the band gap increases. Although in principle α could be influenced by strain induced by the sapphire substrates,[28] the similar structures of $(PEA)_2PbI_4$ and its 2-substituted derivatives suggests that strain would affect the band gaps of $(PEA)_2PbI_4$, 2-F, 2-Cl, and 2-Br similarly. The fact that the materials uniformly, and yet we observe markedly and consistently different values of α suggests strain is not the main cause of band gap changes with temperature.

Figure 1H shows 10 K absorption (solid) and PL (dashed) spectra for $(PEA)_2PbI_4$ (red), 2-F (orange), 2-Cl (green), and 2-Br (blue). As we previously reported,[5,6] the excitonic absorption spectrum of $(PEA)_2PbI_4$ (Figure S7A and Table S3) shows three obvious features separated by 41(2) meV and the PL spectrum (Figure S7B and Table S4) shows hot anti-Stokes PL resonances, corresponding to the resonances observed in absorption. The exciton binding energy is 180(30) meV. Differences in the excitonic substructure for the 2-substituted derivatives (Figure 1H) are immediately evident, with the spacing between resonances being much greater in $(PEA)_2PbI_4$ than in 2-F and 2-Cl and with no obvious excitonic substructure present in 2-Br.

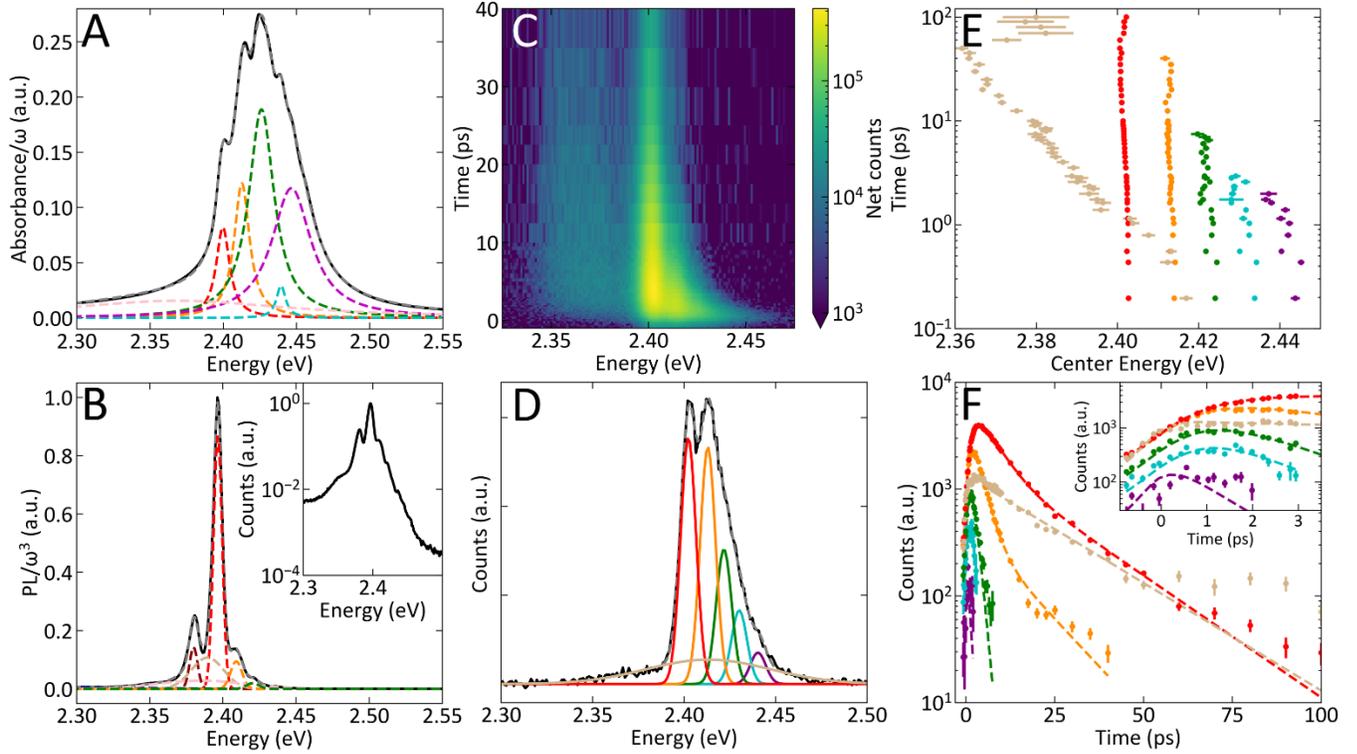

Figure 2: Excitonic (A) absorption and (B) PL spectra (black) of 2-Cl at 10 K. Overall fits are shown in dashed grey with individual resonances in dashed colors. (C) 16 K TRPL spectrum of 2-Cl. (D) Fit of 0.55 ps spectrum from (C). (E) Centers of resonances found from spectrum fits. (F) Areas of each resonance (circles) with fits to kinetic model (dashed lines).

At 10 K, the excitonic absorption spectrum for 2-Cl (Figure 2A) shows four discrete peaks as well as several additional shoulders. We propose a fit of the spectrum to a sum of Lorentzian functions (Table S5). The resonances that correspond to obvious peaks in the absorption spectrum (red, orange, green, blue, Figure 2A) are separated by 13.2(2) meV from one another. There are several additional bumps in the absorption spectrum at higher energies, but to avoid overfitting the spectrum we model the high energy side of the absorption spectrum with a single additional Lorentzian function (magenta, Figure 2A). Second derivative analysis indicates seven inflection points (arrows, Figure S8), suggesting as many as seven discrete resonances. The exciton binding energy is 210(50) meV, 17% larger than in $(PEA)_2PbI_4$.

The 10 K PL spectrum of 2-Cl is fit to the sum of Gaussian functions and shows a 12(1) meV separation between the red, orange, and green resonances (Figure 2B and Table S6), like that observed in absorption. Inset in Figure 2B is the same PL spectrum on a logarithmic scale, which indicates the presence of additional resonances at higher energy that are too small to reliably fit. These additional peaks are also consistently seen in second derivative analysis (arrows, Figure S9). The shift from Lorentzian lineshapes in absorption to Gaussian lineshapes in PL indicates a change from homogeneous to inhomogeneous broadening, which we previously observed in $(PEA)_2PbI_4$.[5] The presence of hot PL and the change in broadening in both $(PEA)_2PbI_4$ and 2-Cl suggest that while the energies of the resonances differ, the dynamics are similar. The Stokes-shifted sharp resonance (dark red, Figure 2B) is separated by 16.5(1) meV from the central red resonance, and the different spacing indicates it may be caused by a distinct phenomenon from the regular progression observed in both the absorption and PL spectra. Interestingly, by changing a single atom on the cation, we increase the fraction of hot PL (orange and green resonances) to 8.4% in 2-Cl from 0.3% in $(PEA)_2PbI_4$.

To investigate the dynamics of the hot carrier PL, we use ultrafast time-resolved PL (TRPL) measurements.[5,29] The TRPL spectrum at 16 K of 2-Cl is shown in Figure 2C. Each slice of the TRPL spectrum is fit to the sum of Gaussian functions (Figure 2D), with the resonance centers shown in Figure 2E and areas in Figure 2F. At early times, we fit a progression consisting of five resonances (Figure 2D), four of which are above the energy of the central resonance at energies consistent with the second derivative analysis of the 10 K steady state PL spectrum (Figure S9). Importantly, the higher energy resonances (purple, cyan, green, orange) decay while the lower energy resonances continue to rise progressively (inset, Figure 2F), like we previously observed in $(PEA)_2PbI_4$.[5] The lifetime of each resonance is quantified by fitting the integrated counts of each resonance versus time using a single or biexponential decay convolved with the measured instrument response function (Figure 2F and Table S7), and with the exception of the broad tan resonance, the lifetime of resonance *m* is used as the risetime for resonance *m-1*. The three highest-energy resonances (purple, cyan, green) have similar singular lifetimes of 0.9(2), 0.9(1), and 1.1(1) ps respectively, and the fourth (orange) resonance exhibits biexponential behavior with lifetimes of 2.4(2) ps and 12(2) ps and a weighted average lifetime of 3.3(4) ps. The central (red) resonance also shows biexponential behavior with lifetimes of 4.5(8) and 19(2) ps and a weighted average lifetime of 8.3(8) ps. The resonances all red-shift in energy over time (Figure 2C), consistent with carrier cooling. There is another resonance (tan) with lifetime 21.5(8) ps that decreases in energy by nearly 30 meV over 200 ps, and this resonance accounts for emission out of other states separate from the observed progression, such as the pink and tan resonances in Figure 2B. The lifetimes of the purple, cyan, and green resonances are similar to those we previously measured for the two hot resonances in $(PEA)_2PbI_4$. The lifetime of the orange resonance is ~3x longer than the lifetime of the hot resonances in $(PEA)_2PbI_4$, which is consistent with a greater fraction of hot PL in 2-Cl than in $(PEA)_2PbI_4$.

Interestingly, the discrete dark red resonance in Figure 2B is missing in TRPL (Figure 3B), and its absence combined with the larger energetic spacing than that which is seen between the other resonances suggests this resonance is not part of the same progression. This resonance takes ns to appear and is seen in 10 K ns TRPL spectra collected by single photon counting (dark red, Figure S10 and Table S8) to exhibit triexponential behavior with lifetimes of 6.5(3) ns, 45(1) ns, and 179(4) ns with a weighted average lifetime of 44(2) ns. The central red resonance is still observed on a ns timescale but is shorter-lived than the dark red resonance, exhibiting biexponential kinetics with lifetimes of 3.1(1) ns and 22(1) ns and a weighted average lifetime of 5.8(3) ns. One possible explanation is that the dark red resonance is caused by an exciton bound to a shallow defect.[30]

We also observe self-trapped exciton resonances (small polarons) in PL centered at energies <2.0 eV that are hypothesized to be the cause of the broad, low-temperature, lower-energy PL resonance present in many 2DHPs including those here in $(PEA)_2PbI_4$, 2-F, 2-Cl, and 2-Br (Figure S11).[31–33] In 2-Cl, the self-trapped exciton resonance is centered at 1.9257(6) eV and exhibits triexponential kinetics with lifetimes of 29.6(2) ns, 301(1) ns, and 3460(20) ns with a weighted average lifetime of 258(2) ns (Figure S12 and Table S9), closer to the timescale of the dark red resonance than the other excitonic PL resonances.

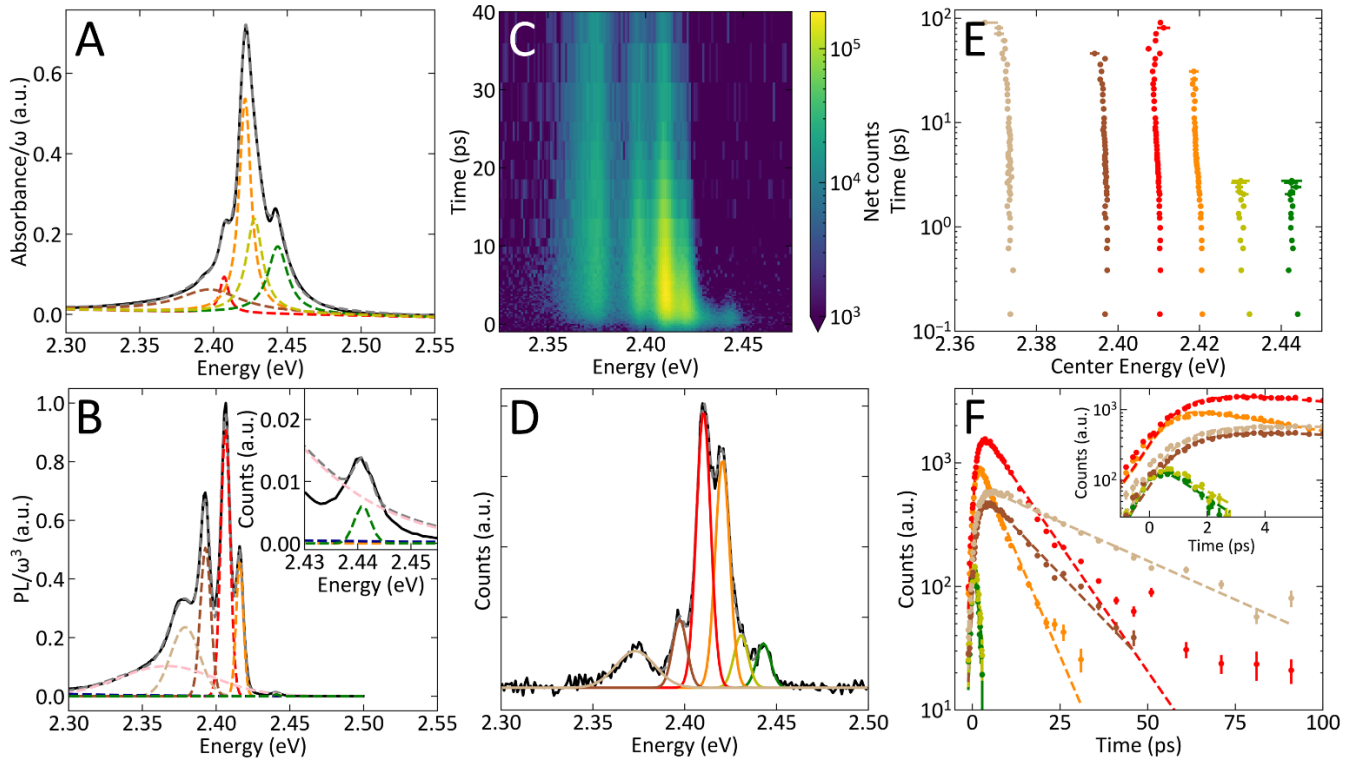

Figure 3: Excitonic (A) absorption and (B) PL spectra (black) of 2-F at 10 K. Overall fits are shown in dashed grey with individual resonances in dashed colors. (C) 15 K TRPL spectrum of 2-F. (D) Fit of 0.62 ps spectrum from (C). (E) Centers of resonances found from spectrum fits. (F) Areas of each resonance (circles) with fits to kinetic model (dashed lines).

Compared to (PEA)$_2$PbI$_4$ and 2-Cl, 2-F exhibits more complex behavior in steady-state and time-resolved spectra. In absorption at 10 K (Figure 3A and Table S10), we observe three sharp Lorentzian resonances (green, orange, red), and another resonance (yellow) of similar width is included for a good fit. There is an inflection point at lower energy than the red resonance indicating the presence of another resonance (brown). Unlike in (PEA)$_2$PbI$_4$ and 2-Cl, there is no regular separation between these resonances. The exciton binding energy is 210(40) meV. Gaussian resonances are used to fit the 10 K PL spectrum (Figure 3B and Table S11). We observe the green, orange, red, and brown resonances, but again there is no consistent spacing and the Stokes shifts are different for each resonance. The yellow resonance is not reliably fit, and two additional broader resonances (tan and pink) are needed for a good fit.

TRPL measurements of 2-F help to clarify the origin of spectral features (Figure 3C). The model used to fit the TRPL lifetimes is identical to the one used for 2-Cl except that since a progression is not obvious, the risetimes of the orange, red, brown, and tan resonances are additional fit parameters. In contrast to the steady-state PL spectrum, the yellow resonance seen in absorption is resolved in TRPL. Unlike in 2-Cl where the hot resonances feed a single sharp lowest-lying resonance in the 12-13 meV progression, in 2-F there are two hot resonances (green and yellow) with lifetimes of 1.6(4) ps and 2.0(5) ps and two lower-energy resonances (red and brown, Figure 3D-F) that have similar risetimes (1.22(4) ps and 1.6(1) ps) and lifetimes (10.6(3) and 14.7(9)) to one another (Table S12). The similarity of these resonances leads us to hypothesize that there are two distinct manifolds of states, with the red and brown resonances being the lowest-lying states in their respective manifold. The two manifolds may originate from the high degree of asymmetry in the Pb-I lattice reflected by the two Pb-I-Pb bond angles differing by almost 3°, or from the additional disorder in the cation compared to (PEA)$_2$PbI$_4$ and 2-Cl. The orange resonance is also present in TRPL with a risetime of 0.38(4) ps and a lifetime of 6.6(3) ps. With dynamics that suggest the presence of two manifolds of states, we are able to identify two sets of regularly spaced resonances in absorption. The first set consists of the green, orange, and brown resonances, which are separated by 22.1(1) and 24(2) meV. The second set consists of the yellow and red resonances, which are separated by 20.4(3) meV. The existence of two manifolds of states would open additional relaxation pathways for hot carriers with the potential for interconversion between the manifolds, which may explain the lack of a direct correspondence between the absorption and PL spectra, such as the unresolved yellow resonance in steady-state PL. If we assume the orange and green resonances in the 10 K steady-state PL spectrum (Figure 2D) are hot based on the kinetics observed in TRPL, they consist of 9.8% of the excitonic PL, 1.4% more than in 2-Cl. Like in 2-Cl, we use single photon counting measurements of 2-F to observe PL and quantify the lifetimes of excitonic resonances on a ns timescale (Figure S13A-C and Table S13). We also observe the self-trapped exciton resonance, which exhibits triexponential behavior and has a weighted average lifetime of 152(1) ns (Figure S13D-F and Table S14).

In 2-Br, we do not see discrete excitonic structure in absorption or PL (Figure 1G and S14A-B and Tables S15-S16), which may be related to the corrugated inorganic framework or to the heavy Br atom reducing the energetic spacing. Like the other species, the absorption spectrum is fit to Lorentzian functions and the PL spectrum to Gaussian functions indicating similar dynamics. The exciton binding energy is 230(30) meV. Even though there are no discrete hot PL resonances in the steady-state spectrum, the TRPL spectrum of 2-Br (Figure S14C-F) shows signatures of hot exciton PL with a resonance (cyan, Figure S14D-F) that has a lifetime of 1.2(3) ps (Table S17). We also observe ns PL in 2-Br out of both the green and orange excitonic resonances (Figure S15) and the self-trapped excitonic resonance (Figure S15 and Tables S18-S19).

These (PEA)$_2$PbI$_4$ derivatives expand the family of 2DHPs that show structure in excitonic absorption and PL spectra, allowing us to test our previous hypothesis that the progressions are vibronic[5] as well as competing hypotheses, most of which involve coupling to phonons. Room-temperature Raman spectra (Figure S16) show vibrational modes near 13 meV, ~25 meV, and 50 meV in (PEA)$_2$PbI$_4$, 2-F, 2-Cl, and 2-Br, indicating there are vibrations at energies consistent with the differences in energy between the observed resonances in absorption and PL spectra.

The existence of many regularly spaced resonances (*e.g.*, > five resonances in 2-Cl) with an energetic spacing that decreases with the mass of the substituent on the cation is consistent with a vibronic progression being responsible for the observed behavior. Vibrational energy levels are evenly spaced, and the kinetics observed in TRPL for (PEA)$_2$PbI$_4$ and 2-Cl indicate that carriers in the excited state can only relax from state *m* to state *m-1* (Figure 3D),[5] which is the vibrational selection rule. We previously hypothesized that a torsional mode on the cation couples to the exciton. This hypothesis is consistent with the smaller energetic spacings in 2-F and 2-Cl and the absence of structure in 2-Br compared to (PEA)$_2$PbI$_4$, because introducing additional off-axis mass would increase the moment of inertia and decrease the energy of a torsional mode.[5,6]

To further investigate the vibrational modes of the cations, we carried out density-functional theory calculations for isolated PEA, 2-F, 2-Cl, and 2-Br cations and analyze the similarity of vibrational modes between the cations. In our previous work, we identified vibrational modes in lead halide 2DHPs containing 4-substituted PEA cations that showed near-perfect similarity to the 411 cm$^{-1}$ (51 meV) mode in the PEA cation. Here, we find a 258 cm$^{-1}$ (32 meV) mode in the 2-F cation, a 164 cm$^{-1}$ (20 meV) mode in the 2-Cl cation, and a 74 cm$^{-1}$ (9 meV) mode in the 2-Br cation that resemble the 411 cm$^{-1}$ (51 meV) mode in the PEA cation and all have energies slightly larger but similar to the energetic spacing between the excitonic resonances (Figure S17). The identified modes in the 2-F, 2-Cl, and 2-Br cations contain additional motion of the $C_2H_4NH_3^+$ backbone compared to the mode in the PEA cation, which may compensate for the motion of the heavier halogen substituents and keep the center-of-mass stationary.

Upon ultrafast photoexcitation in the 3D hybrid perovskite methylammonium lead bromide, transient lattice reconfigurations were observed that took ~1 ps to equilibrate to an excited state nuclear configuration that is distinct from the ground state configuration.[34] This timescale is similar to the lifetimes of the hot PL resonances we observed in (PEA)$_2$PbI$_4$ and see here in 2-F, 2-Cl, and 2-Br.[5] A recent study analyzed coherent phonon oscillations in pump-probe absorption dynamics to demonstrate that a hot resonance and the lowest-lying resonance in the excitonic absorption structure of (PEA)$_2$PbI$_4$ couple differently to low energy (< 8 meV) lattice vibrations, and the authors of this study claim that this observation rules out a vibronic progression in (PEA)$_2$PbI$_4$.[8] We invite further discussion on this topic because: (a) the bleach lifetime of the hot resonance was found to be orders of magnitude longer than the lifetime of the hot excitons (~1 ps),[5] raising the question as to whether the absorptive transients (and observed phononic oscillations) arise from the hot exciton itself or a relaxed state also capable of bleaching the transition; (b) given the observation of transient lattice reconfiguration in the excited state, the ~1 ps lifetime of hot excitons may be too short for these excitons to reach equilibrium in their interaction with the lattice, while longer-lived excitons that reach the lowest-lying excited state will have additional time to equilibrate and may respond to (or generate) lattice modes differently. In addition, the lack of mirror symmetry about the lowest-lying resonance between the excitonic absorption and PL spectra in (PEA)$_2$PbI$_4$, 2-F, or 2-Cl has been used to dispute the vibronic coupling hypothesis.[35] Mirror symmetry would be expected under the Franck-Condon approximation, under which the nuclear coordinates do not change upon electronic excitation,[36] but the observation of different configurations in the ground and excited states in hybrid perovskites is consistent with its absence. Detailed theoretical and experimental studies are necessary to further investigate the nature of the ground and excited states in 2DHPs.

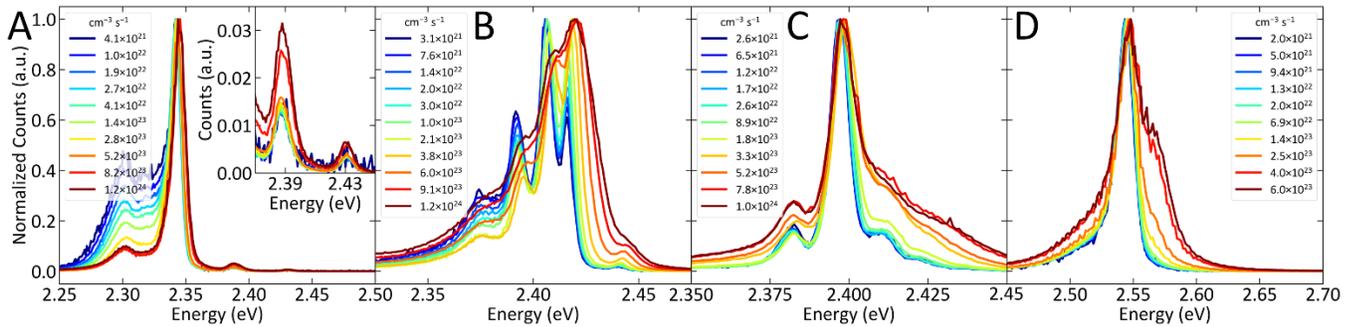

Figure 4: 10 K fluence-dependent PL spectra excited at 3.28 eV normalized to the maximum number of counts at each power of (A) (PEA)$_2$PbI$_4$, (B) 2-F, (C) 2-Cl, and (D) 2-Br.

We probe the power dependence of the hot exciton PL in (PEA)$_2$PbI$_4$, 2-F, 2-Cl, and 2-Br. Using 3.28 eV excitation, we find that at excitation densities $\geq 10^{23}$ photons cm$^{-3}$ s$^{-1}$, the proportion of hot exciton PL increases and there is a small blue-shift of the central PL resonance (Figure S4). One possible explanation for the relative increase in in hot PL is that a hot phonon bottleneck may occur, where a large phonon emission rate in the presence of a high carrier density causes a nonequilibrium phonon population with increased phonon reabsorption and a resulting net decrease in the rate of relaxation to the lowest vibrational level. A similar phonon bottleneck is seen in the 3D hybrid perovskite methylammonium lead iodide at high photoexcitation density using transient absorption spectroscopy.[37] The increased hot PL is consistent with the theory that the resonances are caused by a vibronic progression.

Several other hypotheses have been proposed for the origin of excitonic structure in this class of materials.[8–11] One of these hypotheses is that the excitonic structure is caused by distinct excitonic states likely caused by polaronic effects.[9] Our data here indicate that both vibronic progressions and such distinct excitonic transitions may occur. We note that other than asymmetry in the inorganic framework, which is absent in 2-Cl, no consistent explanation has been given as to why excitonic states of different origin should be regularly spaced in energy and obey the observed kinetics. A modification of the distinct exciton hypothesis is that some of the additional excitonic states are caused by biexcitons combined with the existence of two distinct excitonic states.[11] While biexcitons are likely in 2DHPs, they were observed at excitation densities $>10^{18}/cm^3$. Our steady-state absorption spectra are taken using a benchtop UV-Vis-NIR spectrophotometer resulting in excitation densities $\ll 10^{18}/cm^3$, yet we still observe multiple equally spaced resonances in absorption in $(PEA)_2PbI_4$ and 2-Cl. It is also unclear why the biexciton binding energy would be equal to the spacing between the proposed distinct excitonic states. Lastly, the biexciton binding energy should not change so dramatically between $(PEA)_2PbI_4$, 2-F, and 2-Cl, and 2-Br given the exciton binding energies are within 30% of one another and biexcitons exist in the inorganic lattice.

**Conclusion**

We show single atom substitution on the cation in 2DHPs dramatically alter their structure and photophysical properties seen in steady state and time-resolved absorption and PL spectra. These substitutions can be used to tune the proportion of hot exciton PL, with the amount of hot exciton PL increasing by an order of magnitude to 8.4% in 2-Cl and 9.8% in 2-F from 0.3% in $(PEA)_2PbI_4$. By expanding the family of 2DHPs which show complex structure in excitonic absorption and PL spectra and hot exciton PL, we can evaluate competing theories on their origins. The regular spacing of the excitonic resonances in $(PEA)_2PbI_4$, 2-F, and 2-Cl that inversely correlates with the mass of the off-axis substituent on the cation, and the presence of at least five resonances in 2-Cl, support the hypothesis that a vibronic progression is responsible for the existence of this structure. The kinetics observed in TRPL further support the existence of a vibronic progression because vibrational selection rules are followed. Lastly, hot exciton amplification provides a further route to tune the PL in 2D hybrid perovskites. Our findings highlight the need for further experimental and theoretical studies on the cation's contribution to the optical and electronic properties of 2DHPs because simple changes to the cation have profound effects.

**Methods**

$(PEA)_2PbI_4$, 2-F, 2-Cl, and 2-Br are synthesized by slowly cooling a hot hydriodic acid solution containing perovskite precursors following a previously published procedure.[6] Briefly, unstabilized aqueous hydriodic acid (57% w/w, Sigma-Aldrich) is purified by performing successive liquid-liquid extractions using 10% v/v tributylphosphate (99+%, Acros Organics) in chloroform (Fisher Chemical) until clear in color immediately before use.[6] 231 mg $PbI_2$ (99.999+%, Strem Chemicals) is dissolved in a minimal amount of hydriodic acid at 100 °C under nitrogen. A stoichiometric amount of phenethylamine (>99.5%, Sigma-Aldrich), 2-fluorophenethylamine (97%, Acros Organics), 2-chlorophenethylamine (97%, Acros Organics), or 2-bromophenethylamine (99%, Acros Organics) is added to the flask, and the reaction is cooled to room temperature at a rate of 2-5 °C. Once at room temperature, crystals are harvested by vacuum filtering the reaction mixture, washed with diethyl ether, and dried under vacuum at 50 °C.

CXRD data for 2-Br are collected on a Bruker APEX II CCD area detector using graphite-monochromated Mo-Kα radiation (λ=0.71073Å) at 100 K. SCXRD data for 2-Cl, and 2-F are collected on a Bruker D8 Quest Photon CMOS area detector using graphite-monochromated Mo- Kα radiation (λ=0.71073Å) at 100 K. The strategies for collection are determined by COSMO (Bruker AXS), SAINT (Bruker AXS) performs integration, and SADABS (Bruker AXS) applies scaling and absorption corrections. The initial solution is found using ShelXT[38] and refined using the least-squares method with ShelXL[39] in the OLEX2 GUI.[40]

Thin films are fabricated by dissolving crystals of the perovskites in anhydrous acetonitrile (Sigma-Aldrich) at a concentration of 15 mg/mL and spin-cast on sapphire rounds at 2500 rpm for 20 s in a nitrogen-filled glove box. The films are subsequently annealed in the glove box at 80 °C for 10 min.

Powder diffraction patterns are taken on thin films using a Rigaku SmartLab X-ray diffractometer using Cu-kα radiation operating at 40 kV and 30 mA in θ-2θ geometry.

Absorption and photoluminescence measurements are taken on thin films of the perovskites in an evacuated Advanced Research Systems DE202 cryostat. Absorption spectra are collected in an Agilent Cary 5000 spectrometer. Steady state photoluminescence (PL) and nanosecond time-resolved photoluminescence (TRPL) spectra are collected in an Edinburgh Instruments FLS1000 fluorescence spectrometer using a Hamamatsu R13456 photomultiplier tube detector. Steady state spectra are excited using a monochromated Xe arc lamp, and nanosecond TRPL spectra are excited using a 378 nm Picoquant diode laser. Picosecond TRPL measurements are taken using an optical Kerr-gate setup described previously.[5] Raman spectra are collected on Horiba LabRam HR confocal Raman scattering microscope using a HeNe laser. Steady-state and time-resolved spectra are fit using the lmfit Python module.[41] All reported uncertainties are one standard deviation.

Vibrational modes of isolated cations are calculated using a previously-published method.[6] To compare vibrational modes in different cations, the Kabsch algorithm[42,43] is used to orient the cations, and the atomic displacements of each vibrational modes are projected against one another.

**Acknowledgments**

We thank J. Subotnik for assistance and computational resources for electronic structure calculations. This work is supported by CRK's Stephen J. Angello Professorship. Raman scattering measurements are supported by the National Science Foundation under grant ECCS-1542153 and are performed at the University of Pennsylvania's Singh Center for Nanotechnology, an NNCI member. Ultrafast TRPL measurements are supported by the National Science Foundation under grant DMR-1720530.

**Supporting Information**

Crystal structures of 2-F, 2-Cl, and 2-Br (CIF)

Figures S1-S17, Tables S1-S19 (PDF)

# Supporting Information: Tailoring Hot Exciton Dynamics in 2D Hybrid Perovskites through Cation Modification

Daniel B. Straus[1#], Sebastian Hurtado-Parra[2], Natasha Iotov[2], Qinghua Zhao[1], Michael R. Gau[1], Patrick J. Carroll[1], James M. Kikkawa[2], and Cherie R. Kagan*[134]

Departments of [1]Chemistry, [2]Physics, [3]Electrical and Systems Engineering and [4]Materials Science and Engineering, University of Pennsylvania, Philadelphia, PA 19130

[#]Present address: Department of Chemistry, Princeton University, Princeton, NJ 08544

[*]Author to whom correspondence should be addressed. Email: kagan@seas.upenn.edu

**Additional Figures**

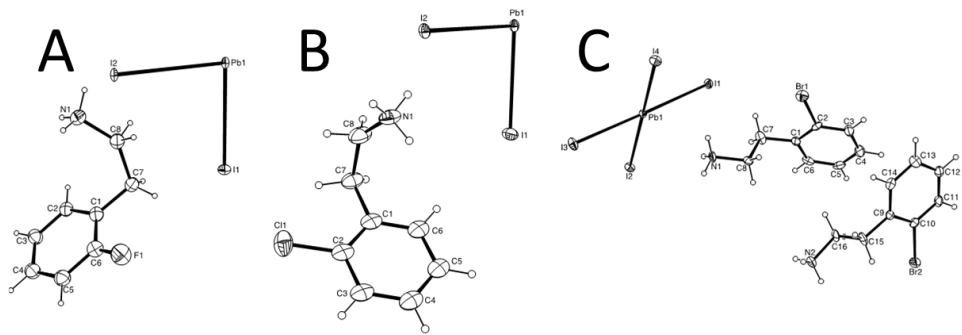

Figure S1: ORTEP drawings of asymmetric units of crystal structures of (A) 2-F, (B) 2-Cl, and (C) 2-Br with 50% thermal ellipsoids. Disorder not shown in (A) and (B).

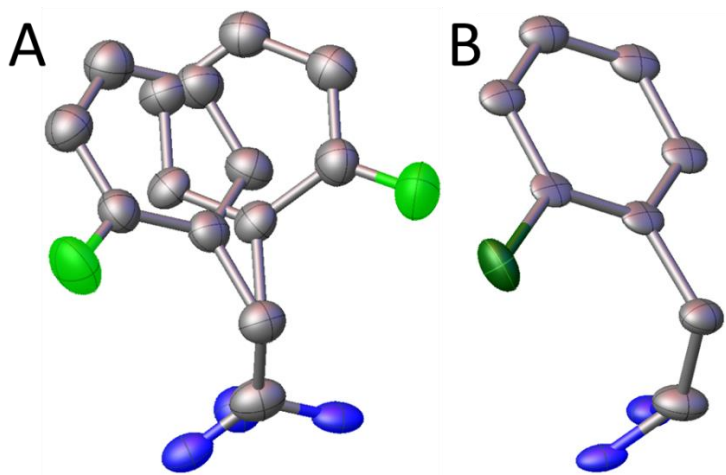

Figure S2: Asymmetric units of cation in (A) 2-F and (B) 2-Cl with hydrogen atoms omitted.

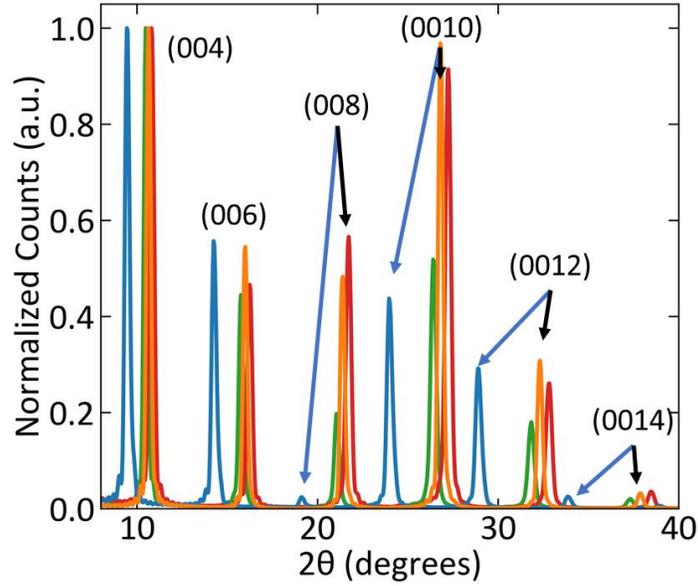

Figure S3: X-ray diffraction patterns of thin films of (red) (PEA)$_2$PbI$_4$, (orange) 2-F, (green) 2-Cl, and (blue) 2-Br. Reflections are indexed according to convention that the *c*-axis is orthogonal to the Pb-I layers.

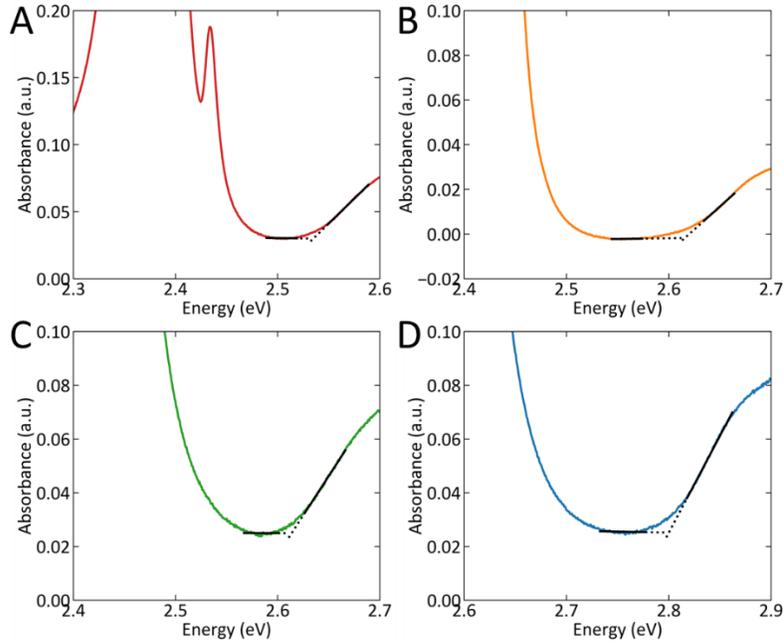

Figure S4: Band gap determination at 10 K using linear fits for (A) (PEA)$_2$PbI$_4$, (B) 2-F, (C) 2-Cl, and (D) 2-Br.

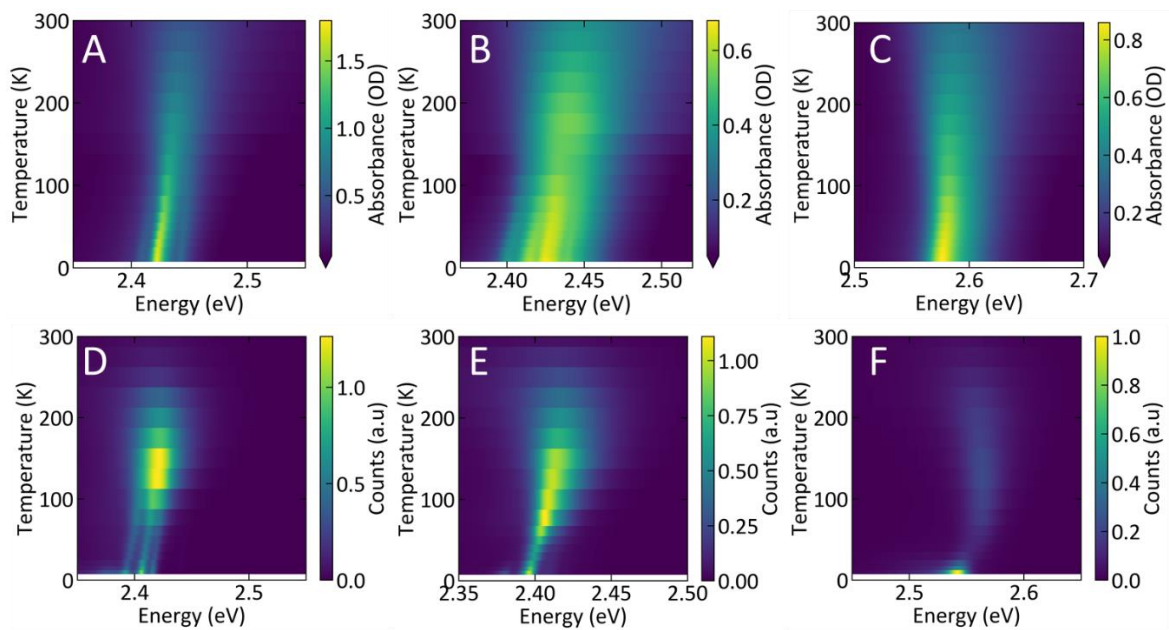

Figure S5: (A-C) Absorption and (D-F) PL spectra excited at 3.10 eV of (A, D) 2-F, (B, E) 2-Cl, and (C, F) 2-Br as a function of temperature.

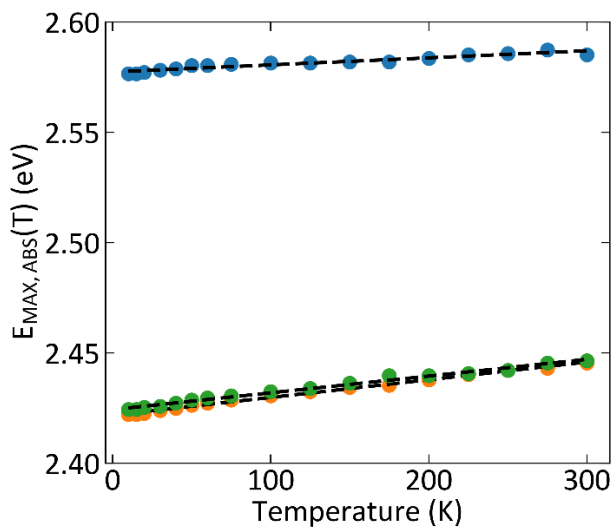

Figure S6: Excitonic absorption maximum versus temperature of 2-F (orange), 2-Cl (green), and 2-Br (blue) with linear fits.

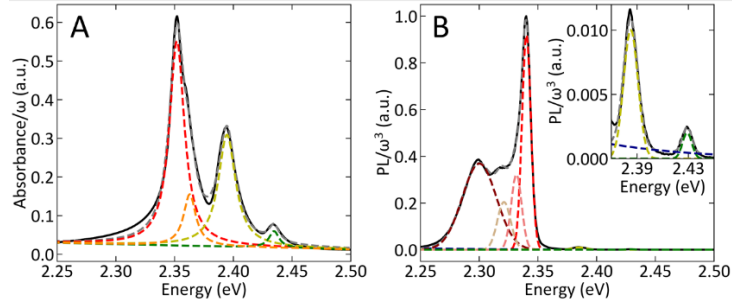

Figure S7: 10 K excitonic absorption and PL spectra (black) of (PEA)$_2$PbI$_4$ with experimental spectrum (black), overall fit (dashed grey), and individual resonance (dashed colors).

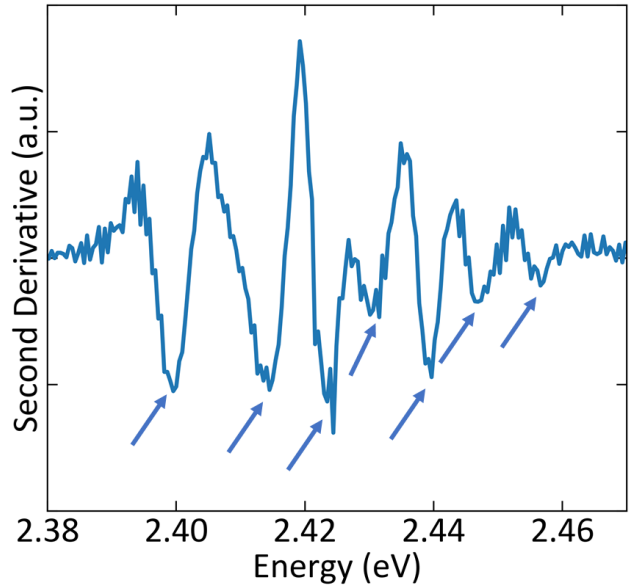

Figure S8: Second derivative of 10 K excitonic absorption spectrum of 2-Cl with inflection points indicated with arrows.

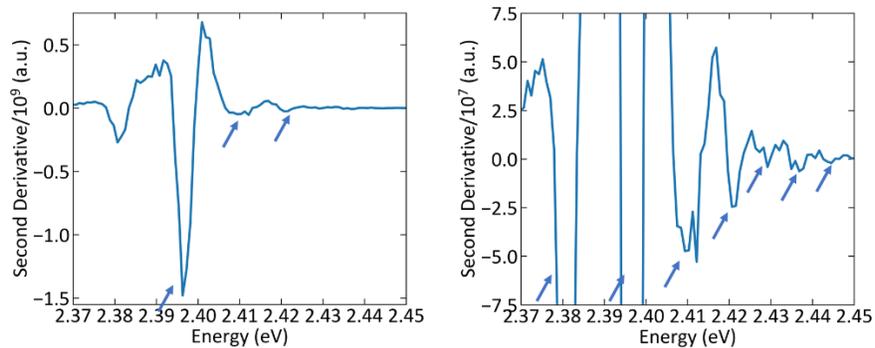

Figure S9: Second derivative of 10 K PL spectrum of 2-Cl with inflection points indicated with arrows.
30

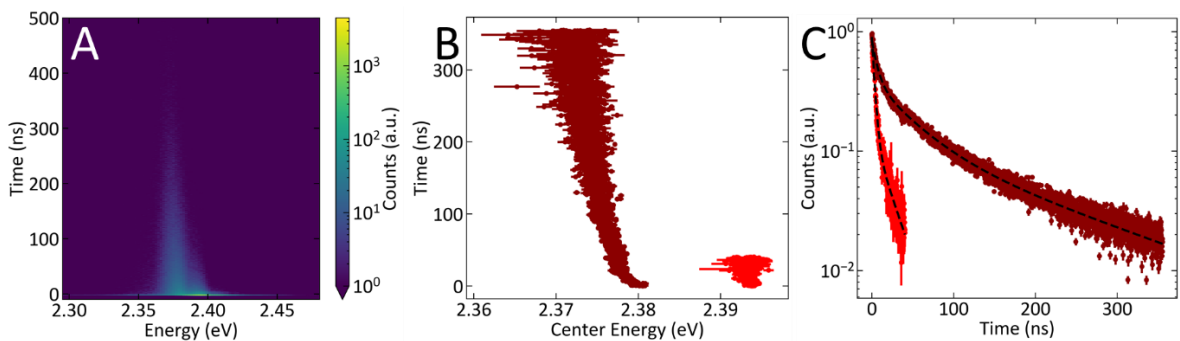

Figure S10: (A) 10 K time-correlated single-photon counting (TCSPC) spectrum of 2-Cl, with (B) resonance centers and (C) lifetime fits.

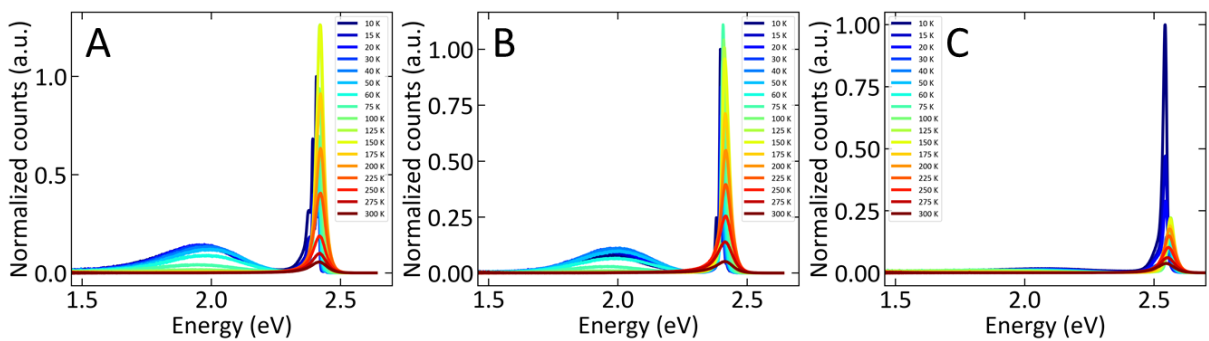

Figure S11: Temperature-dependent PL spectra excited at 3.10 eV for (A) 2-F, (B) 2-Cl, and (C) 2-Br.

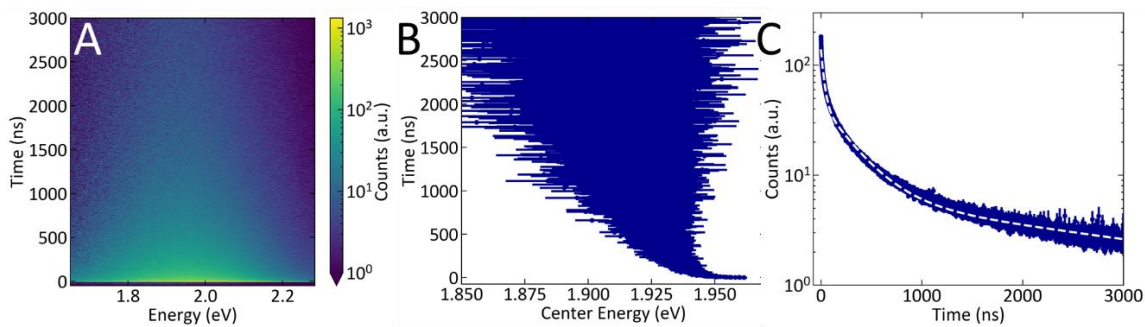

Figure S12: (A) 10 K TCSPC spectrum of the self-trapped excitonic resonance in 2-Cl, with (B) resonance centers and (C) lifetime fit.

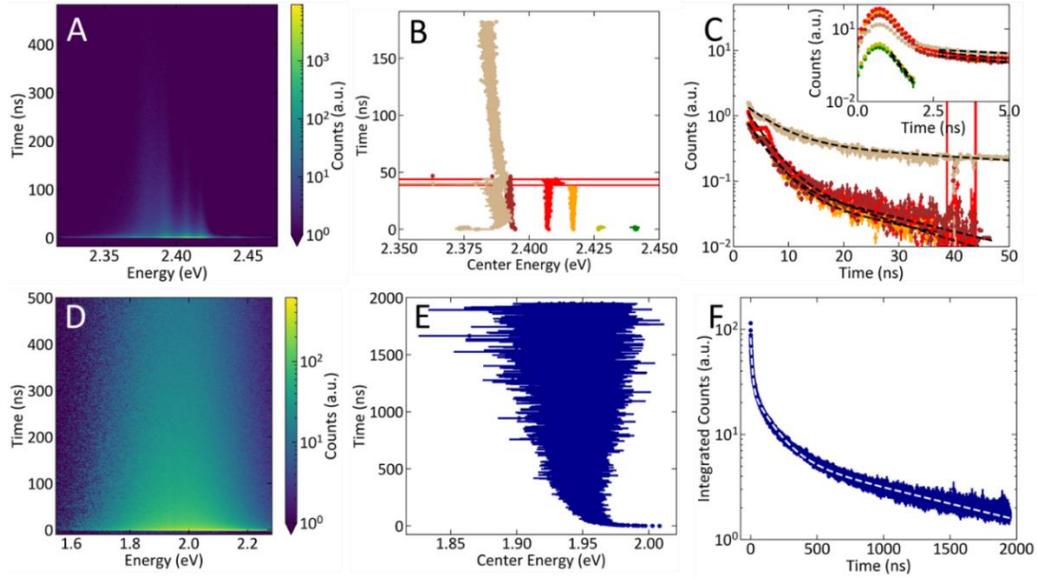

Figure S13: 10 K TCSPC spectra of the 2-F (A) excitonic PL with (B) resonance centers and (C) areas with lifetime fits. The 0.28(1) ns lifetimes of the hot green and yellow resonances are instrumentally limited. The orange, red, brown resonances show biexponential lifetimes and have similar weighted average lifetimes of 4.0(2)-4.3(2) ns. The tan resonance also shows biexponential kinetics with a weighted average lifetime of 30.1(6) ns (Table S13). (D) Self-trapped excitonic PL with (E) and resonance center and (F) area with lifetime fit.

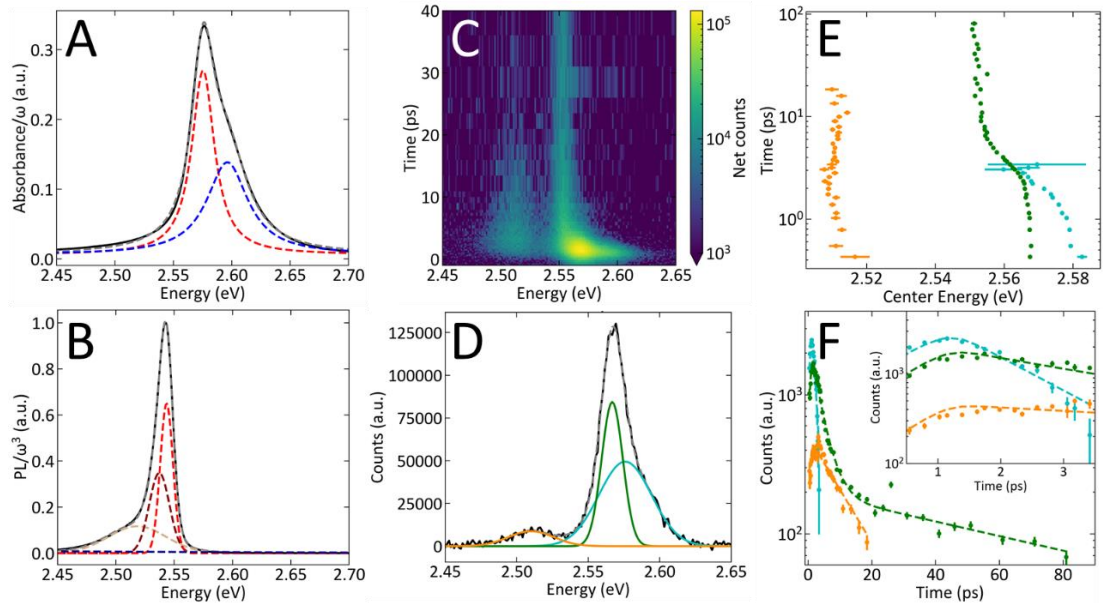

Figure S14: Excitonic (A) absorption and (B) PL spectra (black) of 2-Br at 10 K with overall fit (dashed grey) and individual resonances (dashed colors). (C) 12 K TRPL spectrum of 2-Br, with (D) fit of 1.38 ps spectrum from (C). (E) Centers and (F) areas of fit resonances with lifetime fits (dashed lines). No risetime parameters are used other than convolution with the instrument response function.

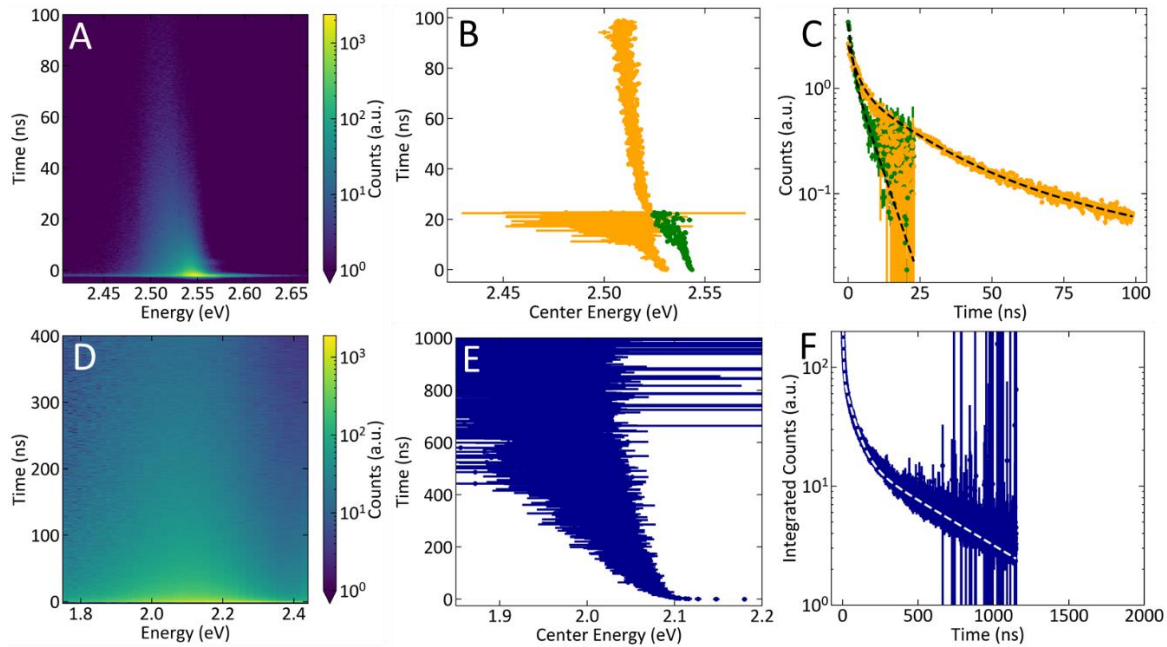

Figure S15: 10 K TCSPC spectra of 2-Br (A) excitonic PL with (B) resonance centers and (C) areas with lifetime fits. (D) Self-trapped excitonic PL with (E) and resonance center and (F) area with lifetime fit.

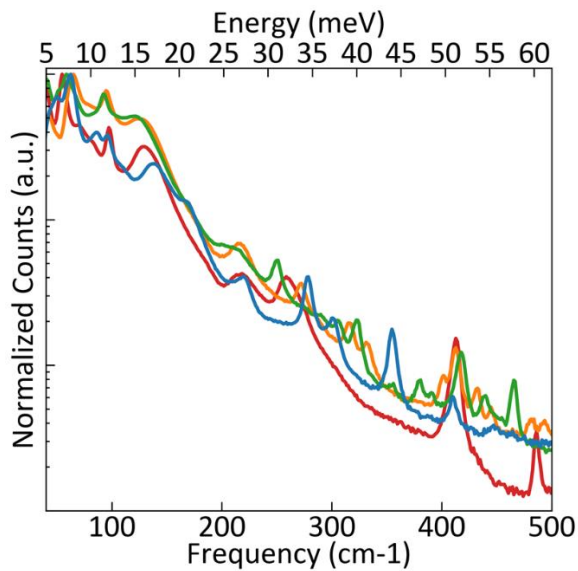

Figure S16: Room temperature Raman spectra of (PEA)$_2$PbI$_4$ (red), 2-F (orange), 2-Cl (green), and 2-Br (blue).

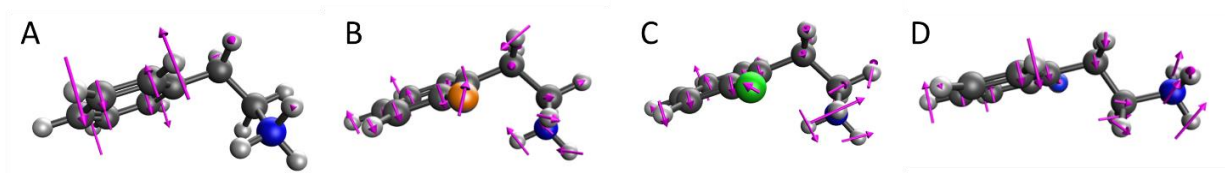

Figure S17: Similar vibrational modes in (A) PEA, (B) 2-F, (C) 2-Cl, and (D) 2-Br.

**Additional Tables**

Table S1: Summary of structural and crystallographic parameters

|  | 2-F | 2-Cl | 2-Br |
| --- | --- | --- | --- |
| **Empirical formula** | $C_{16}H_{22}F_2I_4N_2Pb$ | $C_{16}H_{22}Cl_2I_4N_2Pb$ | $C_{16}H_{22}Br_2I_4N_2Pb$ |
| **Formula weight** | 995.14 | 1028.04 | 1116.96 |
| **Temperature (K)** | 100 | 100 | 100 |
| **Crystal system** | monoclinic | monoclinic | monoclinic |
| **Space group** | C2/m | C2/m | $P2_1$ |
| **a** | 32.8889(14)Å | 33.2779(18)Å | 8.8961(4)Å |
| **b** | 6.1790(3)Å | 6.1691(3)Å | 8.1587(3)Å |
| **c** | 6.0988(3)Å | 6.1839(3)Å | 17.9890(7)Å |
| **β** | 89.989(2)° | 90.255(3)° | 94.847(2)° |
| **Volume** | 1239.40(10)Å$^3$ | 1269.51(11)Å$^3$ | 1300.98(9)Å$^3$ |
| **Z** | 2 | 2 | 2 |
| **Interlayer spacing** | 16.4445(7)Å | 16.6390(9)Å | 17.9890(7)Å |
| **Pb-I-Pb bond angles (deg)** | 151.48(5), 148.73(6) | 151.23(6), 151.21(6) | 139.92(3), 149.17(4) |
| **d$_{calc}$** | 2.667 g/cm$^3$ | 2.689 g/cm$^3$ | 2.851 g/cm$^3$ |
| **μ** | 11.806 mm$^{-1}$ | 11.723 mm$^{-1}$ | 14.306 mm$^{-1}$ |
| **F(000)** | 888 | 920 | 992 |
| **Crystal size, mm** | 0.11 × 0.08 × 0.01 | 0.08 × 0.05 × 0.02 | 0.07 × 0.06 × 0.02 |
| **2θ range for data collection** | 6.68 - 55.002° | 6.588 - 55.122° | 4.544 - 55.068° |
| **Index ranges** | -42 ≤ h ≤ 42, -7 ≤ k ≤ 8, -7 ≤ l ≤ 7 | -42 ≤ h ≤ 41, -8 ≤ k ≤ 7, -7 ≤ l ≤ 7 | -11 ≤ h ≤ 11, -10 ≤ k ≤ 10, -23 ≤ l ≤ 23 |
| **Reflections collected** | 10871 | 9737 | 17743 |
| **Independent reflections** | 1551[R(int) = 0.0610] | 1571[R(int) = 0.0616] | 5992[R(int) = 0.0491] |
| **Data/restraints/parameters** | 1551/260/160 | 1571/129/103 | 5992/103/229 |
| **Goodness-of-fit on F$^2$** | 1.272 | 1.211 | 0.979 |
| **Final R indexes [I>=2σ (I)]** | $R_1$ = 0.0494, $wR_2$ = 0.0866 | $R_1$ = 0.0493, $wR_2$ = 0.0853 | $R_1$ = 0.0324, $wR_2$ = 0.0608 |
| **Final R indexes [all data]** | $R_1$ = 0.0597, $wR_2$ = 0.0904 | $R_1$ = 0.0640, $wR_2$ = 0.0890 | $R_1$ = 0.0419, $wR_2$ = 0.0640 |
| **Largest diff. peak/hole** | 1.51/-1.79 eÅ$^{-3}$ | 1.90/-1.78 eÅ$^{-3}$ | 1.95/-1.03 eÅ$^{-3}$ |
| **Flack parameter** | n/a | n/a | 0.365(8) |

Table S2: Band gaps and exciton binding energies at 10 K.

|  | Band gap (eV) | Exciton binding energy (meV) |
|---|---|---|
| (PEA)2PbI4 | 2.53(3) | 180(30) |
| 2-F | 2.62(4) | 210(40) |
| 2-Cl | 2.61(5) | 210(50) |
| 2-Br | 2.80(3) | 230(30) |

Table S3: Parameters from fit of 10 K absorption spectrum of (PEA)$_2$PbI$_4$. Absent uncertainties signify standard deviations of less than 0.1 meV.

| Resonance | Area (a.u.) | Center (eV) | Sigma (meV) |
|---|---|---|---|
| green | 0.00070(7) | 2.4346(4) | 5.2(7) |
| yellow | 0.0081(1) | 2.3945 | 8.9(1) |
| orange | 0.0026(3) | 2.3628 | 6.3(5) |
| red | 0.0130(3) | 2.3519(1) | 7.8(1) |

Table S4: Parameters from fit of 10 K PL spectrum of (PEA)$_2$PbI$_4$. Absent uncertainties signify standard deviations of less than 0.1 meV.

| Resonance | Area (a.u.) | Center (eV) | Sigma (meV) |
|---|---|---|---|
| green | 0.00002(2) | 2.429(4) | 4(4) |
| yellow | 0.00014(2) | 2.385(1) | 6(1) |
| red | 0.0088(3) | 2.3402 | 3.8 |
| pink | 0.0036(5) | 2.3316(2) | 4.4(3) |
| tan | 0.0029(3) | 2.3210(6) | 5.7(4) |
| dark red | 0.0135(1) | 2.3001(1) | 14.6(1) |
| dark blue | 0.0253(1) | 1.8435(9) | 190(1) |

Table S5: Parameters from fit of 10K absorption spectrum of 2-Cl. Absent uncertainties signify standard deviations of less than 0.1 meV.

| Resonance | Area (a.u.) | Center (eV) | Sigma (meV) |
|---|---|---|---|
| magenta | 0.0060(2) | 2.4465(3) | 16(2) |
| cyan | 0.00035(6) | 2.4394(1) | 3.8(4) |
| green | 0.0063(2) | 2.4261 | 11(2) |
| orange | 0.00244(8) | 2.4127 | 6.4(1) |
| red | 0.00144(3) | 2.3997 | 5.6 |
| pink | 0.0046(1) | 2.368(2) | 94(2) |

Table S6: Parameters from fit of 10K PL spectrum of 2-Cl. Absent uncertainties signify standard deviations of less than 0.1 meV.

| Resonance | Area (a.u.) | Center (eV) | Sigma (meV) |
|---|---|---|---|
| green | 0.0002(1) | 2.420(2) | 4(2) |
| orange | 0.0009(6) | 2.4091(7) | 4(1) |
| red | 0.0064(4) | 2.3965 | 3.0 |
| dark red | 0.0010(1) | 2.3800(1) | 2.7(2) |
| tan | 0.003(1) | 2.390(5) | 11(3) |
| pink | 0.0021(5) | 2.385(4) | 30(3) |
| dark blue | 0.0527(2) | 1.9257(6) | 152.3(6) |

Table S7: Parameters from TRPL lifetime fit of 2-Cl.

| Resonance | Amplitude 1 (a.u.) | $t_1$ (ps) | Amplitude 2 (a.u.) | $t_2$ (ps) |
|---|---|---|---|---|
| purple | 310(50) | 0.9(2) | n/a | n/a |
| cyan | 1500(200) | 0.9(1) | n/a | n/a |
| green | 2900(300) | 1.1(1) | n/a | n/a |
| orange | 4400(300) | 2.4(2) | 400(100) | 12(2) |
| red | 5400(500) | 4.5(8) | 1900(300) | 19(2) |
| tan | 1370(40) | 21.5(8) | n/a | n/a |

Table S8: TCSPC lifetime fit parameters for the excitonic resonances in 2-Cl at 10 K.

| Resonance | Amplitude 1 (a.u.) | $t_1$ (ns) | Amplitude 2 (a.u.) | $t_2$ (ns) | Amplitude 3 (a.u.) | $t_3$ (ns) |
|---|---|---|---|---|---|---|
| red | 0.77(2) | 3.1(1) | 0.13(1) | 22(1) | n/a | n/a |
| dark red | 0.45(2) | 6.5(3) | 0.272(4) | 45(1) | 0.121(5) | 179(4) |

Table S9: TCSPC Lifetime fit parameters for the self-trapped excitonic resonance in 2-Cl at 10 K.

| Amplitude 1 (a.u.) | $t_1$ (ns) | Amplitude 2 (a.u.) | $t_2$ (ns) | Amplitude 3 (a.u.) | $t_3$ (ns) |
|---|---|---|---|---|---|
| 94.5(5) | 29.6(2) | 38.5(2) | 302(1) | 6.22(3) | 3460(20) |

Table S10: Parameters from fit of 10 K absorption spectrum of 2-F. Absent uncertainties signify standard deviations of less than 0.1 meV.

| Resonance | Area (a.u.) | Center (eV) | Sigma (meV) |
|---|---|---|---|
| green | 0.0043(2) | 2.4435(1) | 8.2(2) |
| yellow | 0.0049(5) | 2.4277(3) | 6.7(3) |
| orange | 0.0081(4) | 2.4214 | 4.8 |
| red | 0.00102(9) | 2.4073 | 3.6(2) |
| brown | 0.0047(4) | 2.398(2) | 26(1) |

Table S11: Parameters from fit of 10 K PL spectrum of 2-F. Absent uncertainties signify standard deviations of less than 0.1 meV.

| Resonance | Area (a.u.) | Center (eV) | Sigma (meV) |
|---|---|---|---|
| green | 0.00002(2) | 2.441(1) | 2(1) |
| orange | 0.00279(4) | 2.4160 | 2.4 |
| red | 0.00728(5) | 2.4063 | 3.2 |
| brown | 0.0045(1) | 2.3928 | 3.5 |
| tan | 0.0058(2) | 2.3789(2) | 9.8(2) |
| pink | 0.0083(2) | 2.3666(9) | 32.4(5) |
| dark blue | 0.0958(3) | 1.9139(4) | 145.0(4) |

Table S12: Parameters from TRPL lifetime fit of 2-F.

| Resonance | Amplitude (a.u.) | t (ps) | $t_{rise}$ (ps) |
|---|---|---|---|
| green | 210(30) | 1.6(4) | n/a |
| yellow | 210(30) | 2.0(5) | n/a |
| orange | 1170(20) | 6.6(3) | 0.38(4) |
| red | 2060(30) | 10.6(3) | 1.22(4) |
| brown | 610(20) | 14.7(9) | 1.6(1) |
| tan | 660(20) | 35(2) | 1.3(2) |

Table S13: TCSPC lifetime fit parameters for the excitonic resonances in 2-F at 10 K.

| Resonance | Amplitude 1 (a.u.) | $t_1$ (ns) | Amplitude 2 (a.u.) | $t_2$ (ns) |
|---|---|---|---|---|
| orange | 1.56(6) | 3.4(1) | 0.068(1) | 26(4) |
| red | 2.3(1) | 3.2(1) | 0.14(2) | 16(1) |
| brown | 1.6(1) | 3.0(1) | 0.11(1) | 21(1) |
| tan | 1.21(4) | 7.8(2) | 0.300(2) | 120.3(8) |

Table S14: TCSPC lifetime fit parameters for the self-trapped excitonic resonance in 2-F at 10 K.

| Amplitude 1 (a.u.) | t$_1$ (ns) | Amplitude 2 (a.u.) | t$_2$ (ns) | Amplitude 3 (a.u.) | t$_3$ (ns) |
|---|---|---|---|---|---|
| 56.1(5) | 11.1(1) | 20.8(1) | 148(2) | 6.31(4) | 1420(10) |

Table S15: Parameters from fit of 10K absorption spectrum of 2-Br. Absent uncertainties signify standard deviations of less than 0.1 meV.

| Resonance | Area (a.u.) | Center (eV) | Sigma (meV) |
|---|---|---|---|
| blue | 0.0098(2) | 2.5754 | 11.8(1) |
| red | 0.0089(2) | 2.5962(3) | 21.4(3) |

Table S16: Parameters from fit of 10K PL spectrum of 2-Br. Absent uncertainties signify standard deviations of less than 0.1 meV.

| Resonance | Area (a.u.) | Center (eV) | Sigma (meV) |
|---|---|---|---|
| red | 0.0089(4) | 2.5434 | 5.5 |
| dark red | 0.0076(4) | 2.5372(4) | 8.7(1) |
| tan | 0.0072(1) | 2.5183(5) | 24.8(3) |
| dark blue | 0.0271(2) | 1.875(2) | 358(3) |

Table S17: Parameters from TRPL lifetime fit of 2-Br.

| Resonance | Amplitude 1 (a.u.) | t1 (ps) | Amplitude 2 (a.u.) | t2 (ps) |
|---|---|---|---|---|
| cyan | 5100(800) | 1.2(3) | n/a | n/a |
| orange | 2100(100) | 3.1(3) | 200(20) | 80(20) |
| green | 500(20) | 11(1) | n/a | n/a |

Table S18: TCSPC lifetime fit parameters for the excitonic resonances in 2-Br at 10 K.

| Resonance | Amplitude 1 (a.u.) | t1 (ns) | Amplitude 2 (a.u.) | t2 (ns) | Amplitude 3 (a.u.) | t3 (ns) |
|---|---|---|---|---|---|---|
| green | 2.5(2) | 1.3(2) | 1.6(2) | 5.4(3) | n/a | n/a |
| orange | 1.54(7) | 2.9(2) | 0.84(2) | 19(1) | 0.17(3) | 90(20) |

Table S19: TCSPC lifetime fit parameters for the self-trapped excitonic resonance in 2-Br at 10 K.

| Amplitude 1 (a.u.) | t1 (ns) | Amplitude 2 (a.u.) | t2 (ns) | Amplitude 3 (a.u.) | t3 (ns) |
|---|---|---|---|---|---|
| **316(6)** | 3.82(8) | 70.0(7) | 64.5(8) | 18.2(2) | 574(5) |